\documentclass[journal]{IEEEtran}
\usepackage{graphicx}
\usepackage{lipsum}
\usepackage{color}
\usepackage{amssymb, amsmath}
\usepackage{supertabular}
\usepackage{threeparttable}
\usepackage{setspace}
\usepackage{capt-of}
\usepackage{balance}
\usepackage{cite}
\usepackage{esint}
\usepackage[latin1]{inputenc}
\usepackage{cancel}
\usepackage[ruled]{algorithm2e}
\usepackage{setspace}
\usepackage{array}
\usepackage[none]{hyphenat}
\usepackage{hyphenat}
\ifCLASSINFOpdf
\else
\fi

\begin{document}

\title{Comparison of DCO-OFDM and M-PAM for LED-Based Communication Systems}

\author{{Jie Lian$^\dag$, Mohammad Noshad$^\ddag$  and Ma\"{\i}t\'{e} Brandt-Pearce$^\dag$}
\thanks{This work was funded in part by the National Science Foundation (NSF) through the STTR program, under award number 1521387.}%
\thanks{$^\dag$The authors are with Charles L. Brown Department of Electrical and Computer Engineering, University of Virginia Charlottesville, VA 22904. (Email:jl5qn@virginia.edu;mb-p@virginia.edu)}
\thanks{$^\ddag$The author is with VLNcomm, Charlottesville, VA, 22911. Email:noshad@vlncomm.com}

\thanks{}
\thanks{}
\thanks{}}

\maketitle

\begin{abstract}
   Light-emitting diode (LED)-based communications, such as visible light communications (VLC) and infrared (IR) communications, are candidate techniques to provide short-range and high-speed data transmission. In this paper, $M$-ary pulse amplitude modulation (M-PAM), used as a high bandwidth efficiency scheme, is compared with an optimized DC-biased optical orthogonal frequency division multiplexing (DCO-OFDM) scheme. Considering the bandwidth limit and constrained peak transmitted power characteristics of LEDs, a bit loading algorithm with an optimized modulation index is used for the DCO-OFDM. To reduce the inter-symbol interference caused by LEDs, a waveform design algorithm with a minimum mean squared error (MMSE) equalizer is applied to the M-PAM system. From numerical results, M-PAM with the optimized signal processing can provide a substantially higher data rate than the optimally designed DCO-OFDM for the same performance.
\end{abstract}

\begin{IEEEkeywords}
optical wireless communications, infrared communications, waveform design, MMSE, equalizer, OFDM, bit loading, M-PAM
\end{IEEEkeywords}



\section{Introduction}

\IEEEPARstart L{ight}-emitting diode (LED)-based communications, typical for short-range optical wireless systems, has attracted much attention in recent research due to its many advantages over radio-frequency (RF) communications. By using LEDs as transmitters, visible light communications (VLC) and infrared (IR) communications are immune to RF interference, have low power consumption, low impact on human health, can offer higher security, and can provide potentially high-data-rate transmission. In this paper, we compare two popular modulation schemes often used with LED-based systems: $M$-ary pulse amplitude modulation (M-PAM) and orthogonal frequency division multiplexing (OFDM).

Recently, OFDM has been employed in optical wireless communication (OWC) systems due to its resistance to inter-symbol interference (ISI) and high spectral efficiency \cite{OFDM_Obrien}. Since intensity modulation and direct detection are used in OWC systems, the transmitted signal should be non-negative. Therefore, conventional OFDM cannot be applied directly in OWC. DC-biased optical OFDM (DCO-OFDM) is a popular optical OFDM technique that can be applied in OWC that use incoherent light \cite{DCO_OFDM_CL}. Hermitian symmetric data is used to make the DCO-OFDM signal real. Because of the nonlinear response of LEDs, the DCO-OFDM signal must be clipped, distorting the signal.

Alternatively, pulse-based M-PAM has been explored to yield a ($\mathrm{\log_2}M$)-fold increase in the data rate compared with on-off keying (OOK) \cite{MPAM_PAJO}. However, the transmitted data rate is still limited by the low rise time of LEDs. When the transmitted symbol rate is high, ISI can affect the system performance. Equalization is an effective way to reduce the ISI caused by the low LED bandwidth \cite{post_equ}. Some researchers have discussed pre/post-equalization, software equalization and hardware equalization methods for VLC \cite{ANN,equ_PAM_1,equ_PAM_2,pre_equ,equ_PAM_4}.

In this paper, we compare the performance of DCO-OFDM and M-PAM techniques for OWC systems.
For DCO-OFDM, we model the clipping noise caused by the LEDs' nonlinearity (clipping at both zero and peak current). We optimize the modulation index and the bits loaded on each subcarrier to maximize the transmitted bit rate despite the limited LED bandwidth. For M-PAM, although there is no clipping, the ISI limits the data rate severely. Recently, some researchers proposed a joint waveform design and minimum mean squared error (MMSE) equalization algorithm to combat ISI and multiple access interference in \cite{JOW_ICC}. In this work, the joint optimal waveform design, referred to as JOW, is used to reduce the ISI assuming a single user. We compare the DCO-OFDM with bit loading and M-PAM with JOW for the same bandwidth LED. From numerical results, the M-PAM modulation scheme using the JOW algorithm can support a higher data rate than the optimized DCO-OFDM with the same bit error rate (BER) performance.

A comparison between optical OFDM and PAM was discussed in \cite{Kahn_comparison}. M-PAM with a minimum mean squared error (MMSE) decision feedback equalizer was shown to have a better performance than optical OFDM. However, only zero clipping for optical OFDM is modeled in \cite{Kahn_comparison}, and pre-equalization is not considered.

The remainder of the paper is organized as follows. The optimized DCO-OFDM technique is described in Section~\ref{DCO-OFDM}. In Section~\ref{M-PAM}, we describe the M-PAM system with the JOW algorithm. DCO-OFDM and M-PAM are compared in Section~\ref{numerical_results}. The paper is concluded in Section~\ref{Conclusion}.


\section{Optimized DCO-OFDM} \label{DCO-OFDM}
In this section, we describe how we optimize the DCO-OFDM system. Due to the LED nonlinearity, the signals may be clipped prior to the LED. The optimized DCO-OFDM scheme maximizes the transmitted bit rate by optimizing the modulation index and the bits loaded on all subcarriers.

\begin{figure}
  \centering
  \includegraphics[width=3.4in]{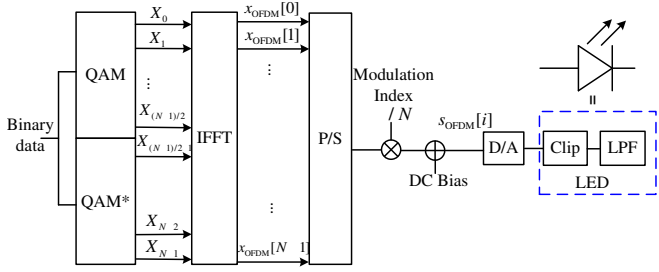}\\
  \caption{Diagram of the DCO-OFDM transmitter with adjustable modulation index.}\label{block_diagram_OFDM}
\end{figure}

A block diagram of the optimized DCO-OFDM is shown in Fig. \ref{block_diagram_OFDM}. To simplify the notation, we analyze the signal in one symbol time. In this diagram, $X_i$ is the data to be modulated on the $i$th subcarrier after $M$-ary quadrature amplitude modulation (QAM). We assume that there are $N$ subcarriers. To make the OFDM signal real, $X_{N-i}$ is the conjugate of $X_i$, $X_{N-i}=X_i^{\ast}$.
After the IFFT, the real OFDM signal for the $k$th component, $x_{\mathrm{OFDM}}[k]$, can be represented as
\begin{equation}\label{ofdm_signal}
   x_{\mathrm{OFDM}}[k]=\sum_{i=0}^{N-1} X_{i}e^{\frac{j2\pi ki}{N}}, ~\forall~k=0,1,\cdots,N-1
\end{equation}
After converting the parallel data to a serial stream and adding a DC offset, the $i$th sample of the transmitted signal can be represented as
   $s_{\mathrm{OFDM}}[i]=\frac{\beta}{N} x_{\mathrm{OFDM}}[i]+s_{\mathrm{dc}}, i=0,1,\cdots,N-1$
where the term $\beta/N$ is referred to as the modulation index. $s_{\mathrm{dc}}$ is the DC bias, which is set to $s_{\mathrm{dc}}=I_{\mathrm{max}}/2$, where $I_{\mathrm{max}}$ is the saturation current to drive the LEDs.

In order to prevent the LEDs from damage, the drive current should remain in the range of $[0,I_{\mathrm{max}}]$. Considering the bandlimited characteristic of LEDs, we model the LED as a clipping component and a lowpass filter in series as shown in Fig. \ref{block_diagram_OFDM}. In reality, the signal outside the range $[0,I_{\mathrm{max}}]$ is clipped. However, the clipping effect can be modeled as a scaling function plus clipping noise.

After matched filtering and sampling at the receiver, the received signal can be modeled as \cite{Jean_optical_OFDM}
\begin{equation}\label{clipped_signal}
\begin{split}
   y_{\mathrm{OFDM}}[i]&=\alpha s_{\mathrm{OFDM}}[i]*h[i]+n_{\mathrm{clip}}[i]+n_{\mathrm{OFDM}}[i], \\
                       &i=0,~1,~\cdots,~N-1
\end{split}
\end{equation}
where $h[i]$ is the discrete time version of the impulse response of the LED. $n_{\mathrm{OFDM}}[i]$ is the additive Gaussian noise with power spectral density $N_o$.
When $N$ is large (usually greater than 64), the analog signal $s_{\mathrm{OFDM}}[i]$ can be modeled as a Gaussian random process. Since the clipping effect is a non-linear operator, the constant coefficient $\alpha$ can be found by using the Bussagang theorem. \cite{Jean_optical_OFDM}
\begin{equation}\label{alpha}
   \alpha=1-\mathrm{erfc}\left(\frac{I_{\mathrm{max}}}{\sqrt{8\sigma_s^2}}\right),
\end{equation}
where $\mathrm{erfc}(x)=2/\sqrt{\pi}\int_{x}^{\infty}e^{-y^2}dy$, and $\sigma_s^2$ is the variance of the discrete-time OFDM signal, $s_{\mathrm{OFDM}}[i]$. We model the clipping noise, $n_{\mathrm{clip}}[i]$, as a zero mean Gaussian variable with a variance estimated using
\begin{equation}\label{clipping_variance}
   \sigma^2_{\mathrm{clip}}=\int_{-\infty}^{0}(\alpha x)^2f(x)dx+\int_{I_{\mathrm{max}}}^{\infty}(\alpha x-I_{\mathrm{max}})^2f(x)dx,
\end{equation}
where $f(\cdot)$ is the probability density function (pdf) of the samples $\alpha s_{\mathrm{OFDM}}[i]$.

In this paper, we assume the 3 dB modulation bandwidth of the LEDs is limited to a few MHz \cite{pre_equ}, and the frequency response can be modeled as a first order lowpass filter. Therefore, the channel amplitude for each subcarrier is not the same. At the receiver, the signal to noise ratio (SNR) for the $i$th subcarrier can be calculated as
\begin{equation}\label{snr_i}
   \gamma_{\mathrm{OFDM}}^{(i)}=\frac{\left(\beta\alpha \mathcal{H}_{i}\right)^2 E\{X_i^2\}}{N\left(\sigma^2_{\mathrm{OFDM}}+\sigma^2_{\mathrm{clip}}\right)},
\end{equation}
where $\mathcal{H}_{i}$ is the LED response for the $i$th subcarrier. $E\{\cdot\}$ represents the expectation operation, and $\sigma_{\mathrm{OFDM}}^2=N_o/T_{\mathrm{OFDM}}$, is the variance of the receiver additive Gaussian noise over one subcarrier. $T_{\mathrm{OFDM}}$ is the symbol time. Given the SNR, we can calculate the bit error rate (BER) for each subcarrier by using the approximate expression \cite{OWC_book}
\begin{equation}\label{ber_i}
  \!\! \mathrm{BER}_i\!\approx\!\!\frac{\sqrt{\!\!M_{\mathrm{OFDM}}^{(i)}}\!-\!1}{\sqrt{\!\!M_{\mathrm{OFDM}}^{(i)}}\log_2\!\!\left(\!\!\sqrt{\!\!M_{\mathrm{OFDM}}^{(i)}}\!\right)}\mathrm{erfc}\!\!\left(\!\!\sqrt{\frac{3\gamma_{\mathrm{OFDM}}^{(i)}}{2M_{\mathrm{OFDM}}^{(i)}\!-\!2}}\right)\!,\! ~\!\forall\!~\!i,
\end{equation}
where $M_{\mathrm{OFDM}}^{(i)}$ is the modulation constellation size for the QAM used in the $i$th subcarrier.

The throughput for the DCO-OFDM can be calculated as
\begin{equation}\label{r_b}
   R_{b}=\frac{1}{T_{\mathrm{OFDM}}}\sum_{i=0}^{(N-1)/2}\log_2M_{\mathrm{OFDM}}^{(i)},
\end{equation}
To optimize the throughput, we can choose the optimal $T_{\mathrm{OFDM}}$, $\beta$, and the number of bits loaded onto each subcarrier. In this paper, for each subcarrier, the subcarrier bit loading is constrained only by the BER requirement, $B^{\mathrm{max}}$.

\section{Optimized M-PAM} \label{M-PAM}
In this section, an optimized M-PAM algorithm using a temporal waveform design at the transmitter and MMSE equalization at the receiver is described. This joint optimal waveform (JOW) design algorithm is based on a pre-coding technique that was proposed recently in \cite{JOW_ICC}. In JOW, the waveforms and MMSE filters can be optimally designed to reduce ISI and support high-speed data transmission rates.

\begin{figure}
  \centering
  \includegraphics[width=3.0in]{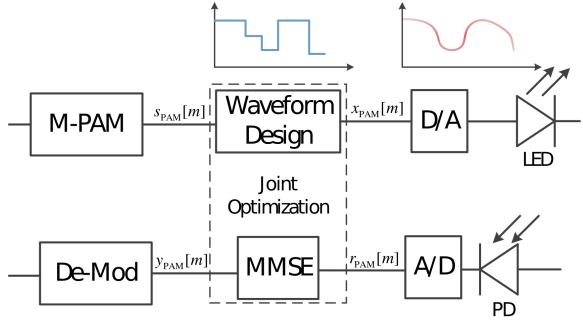}\\
  \caption{Block diagram of JOW using adaptive M-PAM.}\label{JOW_diagram}
\end{figure}

A block diagram of the system is shown in Fig. \ref{JOW_diagram}. $s_{\mathrm{PAM}}[m]\in \{0, \frac{1}{M_{\mathrm{PAM}}-1}, \frac{2}{M_{\mathrm{PAM}}-1}, \cdots, 1\}$ is the $m$th data after M-PAM, where $M_{\mathrm{PAM}}$ is the modulation constellation size. The discrete time transmitted signal after waveform design can be represented as
\begin{equation}\label{transmit_signal}
   x_{\mathrm{PAM}}[i]=\sum_{m=-\infty}^{\infty}s_{\mathrm{PAM}}[m]f[i-mL_f],
\end{equation}
where $\mathbf{f}=(f[1], f[2], \cdots, f[L_f])^T$ is the designed waveform, and $L_f$ is the number of samples describing the waveform, which is a design parameter. Considering the peak drive current constraint, $f[i],~\forall~i$, should be in the range $[0, I_{\mathrm{max}}]$. We assume that the sampling rate, $R_c$, for the waveform design and at the receiver are the same.

After rectangular matched filtering and sampling, the received signals can be represented as
\begin{equation}\label{received_signal}
         r_{\mathrm{PAM}}[i]=\sum_{k=1}^{L_h}x_{\mathrm{PAM}}[i-k]h[k]+n_{\mathrm{PAM}}[i],
\end{equation}
where $\mathbf{h}=(h[1], h[2], \cdots, h[L_h])^T$, is the discrete time version of the truncated impulse response of the LED. $n_{\mathrm{PAM}}[m]$ is the additive receiver noise. After applying the MMSE filter, $\mathbf{w}=(w[1], w[2], \cdots, w[L_w])^T$, the estimated data can be written in matrix form as
\begin{equation}\label{y}
   y_{\mathrm{PAM}}[m]=\mathbf{w}^T \mathbf{H} \mathbf{x}+\mathbf{w}^T\mathbf{n}_{\mathrm{PAM}}+b,
\end{equation}
where $\mathbf{H}$ is a Toeplitz matrix that can be represented as $\mathbf{H}=\left(S_L(\mathbf{h},\frac{L_w-1}{2}),\cdots,\mathbf{h},\cdots,S_R(\mathbf{h},\frac{L_w-1}{2})\right)^T$, where $S_L(\mathbf{h},m)$ and $S_R(\mathbf{h},m)$ are functions that operate as $m$ circular shifts of $\mathbf{h}$ to the left and right, respectively. The vector of transmitted samples that affect $y_{\mathrm{PAM}}[m]$ is denoted,
$\mathbf{x}_{\mathrm{PAM}}=(x_{\mathrm{PAM}}[-N_u],\cdots,x_{\mathrm{PAM}}[0],\cdots,x_{\mathrm{PAM}}[N_l])^T$, where $N_l+N_u+1=L_h$, which describes the length of successive samples that blur together. $N_l$ and $N_u$ represent past and future samples that contribute to ISI, respectively.

The MMSE filter can be obtained as
\begin{equation}\label{MMSE_filter}
   \mathbf{w}=(\mathbf{V}+\sigma_{\mathrm{PAM}}^2\mathbf{I})^{-1}\mathbf{U}, \quad b=\frac{1}{2}-\mathbf{w}^T\mathbf{H}E\left\{\mathbf{x}_{\mathrm{PAM}}\right\}
\end{equation}
where $\mathbf{V}=\mathbf{H} \mathbf{\Sigma} \mathbf{H}$, and $\mathbf{U}=\mathbf{H}E\left\{s_{\mathrm{PAM}}[i]\cdot\mathbf{x}_{\mathrm{PAM}}\right\}$. The additive constant term $b$ is needed for filtering the nonzero-mean signal. $\mathbf{I}$ is the identity matrix, and $\sigma_{\mathrm{PAM}}^2$ is the noise variance that can be calculated as $\sigma_{\mathrm{PAM}}^2=N_oR_c$, where $N_o$ and $R_c$ are noise spectral density and sampling rate, respectively \cite{JOW_ICC}. $\mathbf{\Sigma}$ can be calculated as $\mathbf{\Sigma}=E\{\mathbf{x}_{\mathrm{PAM}}\mathbf{x}_{\mathrm{PAM}}^T\}$, with the $\left((i-1)N_s+u, (j-1)N_s+v\right)$th element of $\mathbf{\Sigma}$ given by
\begin{equation}\label{Sigma_x_element}
  \sigma_{ijuv}^2= \left\{
  \begin{array}{l l}
    \begin{split}
        &\frac{2M_{\mathrm{PAM}}^2-M_{\mathrm{PAM}}}{6M_{\mathrm{PAM}}-6}f[u]f[v]
    \end{split} &, \quad i=j\\
    \frac{1}{4} f[u]f[v] &, \quad i\neq j
  \end{array} \right..
\end{equation}
After the MMSE filter, the power of the ISI plus noise can be estimated as
\begin{equation}\label{ISI_noise}
   \begin{split}
        \sigma^2_{\mathrm{ISI,noise}}&=E\left\{(y_{\mathrm{PAM}}[m]-s_{\mathrm{PAM}}[m])^2\right\} \\
                                     &=\mathbf{w}^T\mathbf{H}\mathbf{\Sigma} \mathbf{H}^T \mathbf{w} + 2b\mathbf{w}^T\mathbf{H}E\left\{\mathbf{x}_{\mathrm{PAM}}\right\}\\
                                     &-2\mathbf{w}^T\mathbf{H}E\left\{s_{\mathrm{PAM}}[i]\cdot\mathbf{x}_{\mathrm{PAM}}\right\}+\left(\mathbf{w}^T\mathbf{H}E\left\{\mathbf{x}_{\mathrm{PAM}}\right\}\right)^2
   \end{split}
\end{equation}
Thus, the signal to interference plus noise ratio (SINR) can be calculated as \cite{Digi_comm_book}
\begin{equation}\label{SINR_k}
  \gamma_{\mathrm{PAM}}=\frac{2M_{\mathrm{PAM}}^2-M_{\mathrm{PAM}}}{(6M_{\mathrm{PAM}}-6)\sigma^2_{\mathrm{ISI,noise}}}
\end{equation}
After substituting (\ref{MMSE_filter}) and (\ref{ISI_noise}) into (\ref{SINR_k}), we can find that $\mathbf{f}$, $M_{\mathrm{PAM}}$ and $R_c$ are the only variables needed to find $\gamma_{\mathrm{PAM}}$. Denoting the SINR as $\gamma_{\mathrm{PAM}}(\mathbf{f},M_{\mathrm{PAM}},R_c)$, the BER can be approximately calculated as \cite{OWC_book}.
\begin{equation}\label{BER_M_PAM}
  \!\!\!\!\mathrm{BER}\!\approx\!\frac{M_{\mathrm{PAM}}-1}{M_{\mathrm{PAM}}\mathrm{log_2}M_{\mathrm{PAM}}}\mathrm{erfc}\!\left(\!\!\sqrt{\frac{\gamma_{\mathrm{PAM}}(\mathbf{f},M_{\mathrm{PAM}},R_c)}{(M_{\mathrm{PAM}}-1)^2}}\right)
\end{equation}

For different sampling rates and modulation constellation sizes, the waveform design algorithm can adaptively adjust the waveforms to minimize the ISI. For a given sampling chip rate and modulation constellation size, the optimized waveforms can be obtained by maximizing the SINR. The optimization cost function is
\begin{equation}\label{opt_func}
   \mathbf{f}^{\ast}=\arg \max_{\mathbf{f}} \gamma_{\mathrm{PAM}}(\mathbf{f},M_{\mathrm{PAM}},R_c),
\end{equation}
where $\mathbf{f}^{\ast}$ is the optimal value for $\mathbf{f}$.

Since the transmission bit rate can be calculated as $R_b=R_c\log_2M_{\mathrm{PAM}}/L_f$,
we need to solve the following problem to maximize the bit rate:
\begin{equation}\label{max_R_b}
\begin{split}
   &\max_{M_{\mathrm{PAM}}}~R_b \\
   s.t.&\frac{M_{\mathrm{PAM}}-1}{M_{\mathrm{PAM}} \log_2M_{\mathrm{PAM}}} \mathrm{erfc}\!\!\left( \!\!\sqrt{\frac{\gamma_{\mathrm{PAM}}(\mathbf{f}^{\ast}, M_{\mathrm{PAM}} ,R_c)}{(M_{\mathrm{PAM}}-1)^2}}\right)\!\!<\!\!B^{\mathrm{max}},
\end{split}
\end{equation}
where $B^{\mathrm{max}}$ is the required BER, and $\mathbf{f}^{\ast}$ can be solved using (\ref{opt_func}) for a given $M_{\mathrm{PAM}}$ and $R_c$.

%
\section{Simulation Results and Comparison} \label{numerical_results}

In this section, numerical results of the comparison between DCO-OFDM and M-PAM are shown. To obtain a fair comparison, the same parameters are used for DCO-OFDM and M-PAM. Unless otherwise noted, the parameters used to obtain the numerical results are shown in Table~\ref{para}. To simplify the problem, an ideal channel (zero loss) is considered in this paper. In addition, the forward current to optical power conversion ratio of the LED is assumed to be unity. Thus, the LED drive current constraint imposes a constraint on the peak transmitted optical power.

\begin{table}
\caption{Parameters Used for Numerical Results}
\label{para}
{\small \begin{center}
  \begin{tabular}{l l}
    \hline
    3 dB bandwidth of LEDs, $f_{3dB}$ & 20~MHz \\
    Number of subcarriers, $N$ & 64 \\
    Noise spectral density, $N_0$ & $3\times10^{-9}$ mW/Hz \\
    BER requirement, $B^{\mathrm{max}}$ & $10^{-3}$\\
    \hline
  \end{tabular}
\end{center}}
\end{table}

\begin{figure}
  \centering
  \includegraphics[width=3.1in]{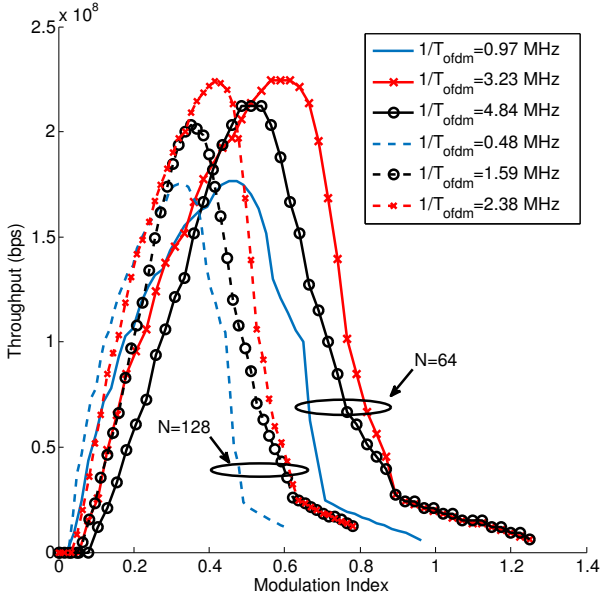}\\
  \caption{Throughput of DCO-OFDM with bit loading for different modulation indexes. The peak transmitted power is $10~\mathrm{mW}$.}\label{bit_loading}
\end{figure}

For DCO-OFDM, a compromise is reached between the signal power and clipping noise power by adjusting the modulation index. In Fig. \ref{bit_loading}, the throughput with different modulation indexes using bit loading is shown. From the results, the number of subcarriers does not seem to affect the maximum throughput.

\begin{figure}
  \centering
  \includegraphics[width=3.0in]{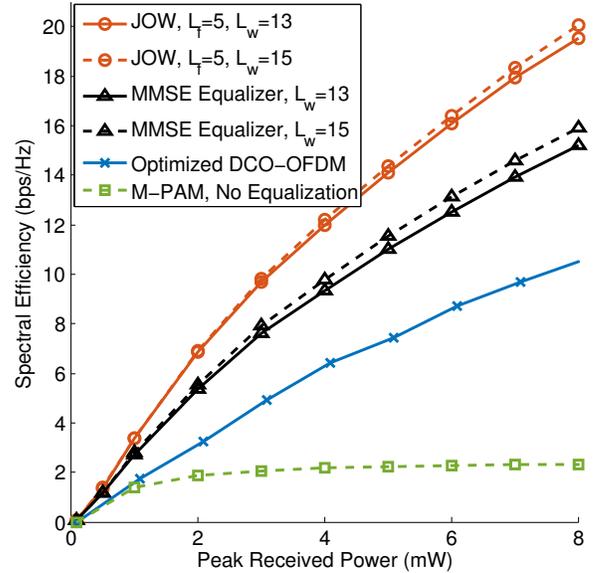}\\
  \caption{Throughput comparison of optimized DCO-OFDM and optimized M-PAM for a single user.}\label{throughput_comparison}
\end{figure}

Fig. \ref{throughput_comparison} shows a comparison between the optimized DCO-OFDM and the optimized M-PAM using JOW. In this figure, $R_b/f_{\mathrm{3dB}}$ is used to measure the spectral efficiency. M-PAM and DCO-OFDM can both use more than the 3 dB bandwidth of the LED. However, the clipping distortion caused by the nonlinearity of the LED can affect the performance of the DCO-OFDM; therefore, using M-PAM with JOW can provide a better performance than DCO-OFDM. From the results, the M-PAM using JOW can support an $80\%$ higher transmitted bit rate than the optimized DCO-OFDM for the parameters tested. With the help of waveform design, JOW can also surpass M-PAM that uses only the MMSE equalizer. If there is no equalization technique for M-PAM, the optimized DCO-OFDM can support about five times higher data rate than M-PAM when the transmitted power is 8~mW.

%
%
\section{Conclusion} \label{Conclusion}
In this paper, we compare the performance of DCO-OFDM and M-PAM for LED-based communication systems. We consider the bandlimited characteristic and the constrained transmitted power of LEDs. We propose an optimized DCO-OFDM by choosing the modulation index that maximizes the data rate with bit loading. For M-PAM, a joint optimization of the waveform and the receiver filter is proposed to reduce the ISI caused by the bandlimited LED. From numerical results, we conclude that M-PAM with waveform design and MMSE equalization can provide a significantly higher data transmission rate than DCO-OFDM with bit loading and an optimal modulation index.

\bibliographystyle{IEEEtran}
\bibliography{myref}

\begin{thebibliography}{10}
\providecommand{\url}[1]{#1}
\csname url@samestyle\endcsname
\providecommand{\newblock}{\relax}
\providecommand{\bibinfo}[2]{#2}
\providecommand{\BIBentrySTDinterwordspacing}{\spaceskip=0pt\relax}
\providecommand{\BIBentryALTinterwordstretchfactor}{4}
\providecommand{\BIBentryALTinterwordspacing}{\spaceskip=\fontdimen2\font plus
\BIBentryALTinterwordstretchfactor\fontdimen3\font minus
  \fontdimen4\font\relax}
\providecommand{\BIBforeignlanguage}[2]{{%
\expandafter\ifx\csname l@#1\endcsname\relax
\typeout{** WARNING: IEEEtran.bst: No hyphenation pattern has been}%
\typeout{** loaded for the language `#1'. Using the pattern for}%
\typeout{** the default language instead.}%
\else
\language=\csname l@#1\endcsname
\fi
#2}}
\providecommand{\BIBdecl}{\relax}
\BIBdecl

\bibitem{OFDM_Obrien}
A.~Azhar, T.~Tran, and D.~O'Brien, ``A {Gigabit/s} indoor wireless transmission
  using {MIMO-OFDM} visible-light communications,'' \emph{IEEE Photon. Technol.
  Lett.}, vol.~25, no.~2, pp. 171--174, 2013.

\bibitem{DCO_OFDM_CL}
M.~Zhang and Z.~Zhang, ``An optimum {DC}-biasing for {DCO-OFDM} system,''
  \emph{IEEE Commun. Lett.}, vol.~18, no.~8, pp. 1351--1354, Aug 2014.

\bibitem{MPAM_PAJO}
J.~Lian and M.~Brandt-Pearce, ``Adaptive {M-PAM} for multiuser {MISO} indoor
  {VLC} systems,'' in \emph{2016 IEEE Global Commun. Conf. (GLOBECOM)}, 2016,
  pp. 1--7.

\bibitem{post_equ}
H.~Li, X.~Chen, B.~Huang, D.~Tang, and H.~Chen, ``High bandwidth visible light
  communications based on a post-equalization circuit,'' \emph{IEEE Photon.
  Technol. Lett.}, vol.~26, no.~2, pp. 119--122, 2014.

\bibitem{ANN}
P.~Haigh, Z.~Ghassemlooy, S.~Rajbhandari, I.~Papakonstantinou, and W.~Popoola,
  ``Visible light communications: 170 {Mb/s} using an artificial neural network
  equalizer in a low bandwidth white light configuration,'' \emph{J. Lightw.
  Technol.}, vol.~32, no.~9, pp. 1807--1813, 2014.

\bibitem{equ_PAM_1}
I.~N. Osahon, E.~Pikasis, S.~Rajbhandari, and W.~O. Popoola, ``Hybrid {POF/VLC}
  link with {M-PAM} and {MLP} equaliser,'' in \emph{2017 IEEE International
  Conf. on Commun. (ICC)}, May 2017, pp. 1--6.

\bibitem{equ_PAM_2}
X.~Li, N.~Bamiedakis, X.~Guo, J.~J.~D. McKendry, E.~Xie, R.~Ferreira, E.~Gu,
  M.~D. Dawson, R.~V. Penty, and I.~H. White, ``Wireless visible light
  communications employing feed-forward pre-equalization and {PAM}-4
  modulation,'' \emph{J. Lightw. Technol.}, vol.~34, no.~8, pp. 2049--2055,
  April 2016.

\bibitem{pre_equ}
X.~Huang, J.~Shi, J.~Li, Y.~Wang, and N.~Chi, ``A {Gb/s} {VLC} transmission
  using hardware preequalization circuit,'' \emph{IEEE Photon. Technol. Lett.},
  vol.~27, no.~18, pp. 1915--1918, 2015.

\bibitem{equ_PAM_4}
L.~Grobe and K.~D. Langer, ``Block-based {PAM} with frequency domain
  equalization in visible light communications,'' in \emph{2013 IEEE Globecom
  Workshops (GC Wkshps)}, Dec 2013, pp. 1070--1075.

\bibitem{JOW_ICC}
J.~Lian and M.~Brandt-Pearce, ``Joint optimal waveform design for multiuser
  {VLC} systems over {ISI} channel,'' in \emph{2016 IEEE International Conf. on
  Commun.}, May 2017, pp. 1--6.

\bibitem{Kahn_comparison}
D.~J.~F. Barros, S.~K. Wilson, and J.~M. Kahn, ``Comparison of orthogonal
  frequency-division multiplexing and pulse-amplitude modulation in indoor
  optical wireless links,'' \emph{IEEE Trans. on Commun.}, vol.~60, no.~1, pp.
  153--163, January 2012.

\bibitem{Jean_optical_OFDM}
J.~Armstrong, ``{OFDM} for optical communications,'' \emph{J. Lightw.
  Technol.}, vol.~27, no.~3, pp. 189--204, Feb 2009.

\bibitem{OWC_book}
Z.~Ghassemlooy, W.~Popoola, and Rajbhandari, \emph{Optical Wireless
  Communications: System and Channel Modeling with MATLAB}.\hskip 1em plus
  0.5em minus 0.4em\relax CRC Press, 2013.

\bibitem{Digi_comm_book}
J.~G. Proakis, \emph{Digital Communications, 5th edition}.\hskip 1em plus 0.5em
  minus 0.4em\relax McGraw-Hill Higher Education, 2008.

\end{thebibliography}

\end{document}